\documentclass[letter,longauth]{aa}  

\usepackage{graphicx}
\usepackage{txfonts}
\usepackage{CJKutf8} 
\usepackage{textcomp} 

\usepackage{xspace}
\usepackage[dvipsnames]{xcolor}
\usepackage[colorlinks=true]{hyperref}

\hypersetup{
    colorlinks=true,
    linkcolor=blue,
    filecolor=blue,      
    urlcolor=blue,
    citecolor=teal
}

\newcommand{\prospector}{\textsc{Prospector}\xspace}

\newcommand{\Mstar}{\ensuremath{\mathrm{M}_\star}\xspace}
\newcommand{\src}{Leviathan}   
\newcommand{\Msun}{\ensuremath{\mathrm{M}_\odot}\xspace}

\newcommand{\ha}{H\,$\alpha$}

\newcommand{\Msrc}{\ensuremath{\mathrm{\Mstar=\rm 10^{11.34}M}_\odot}\xspace}

\newcommand{\zh}[1]{\begin{CJK}{UTF8}{bsmi}#1\end{CJK}}

\begin{document}

   \title{
   Leviathan: a very massive, extended sub-millimetre galaxy spectroscopically confirmed at $\rm z = 4.05$
   }
   \titlerunning{\src: a massive, extended SMG at $z=4$}

   \author{William M. Baker
          \inst{1}\fnmsep\thanks{william.baker@nbi.ku.dk}
          \and
          Francesco Valentino\inst{2,3}
          \and
          Jens Hjorth\inst{1}
          \and
          Gabriel Brammer\inst{2,4}
          \and 
          Ugn\.{e} Dudzevi\v{c}i\={u}t\.{e}\inst{2,3}
          \and
          Kei Ito\inst{2,3}
          \and
          Antonio Pensabene\inst{2,3}
          \and
          Pengpei Zhu (\zh{朱芃佩})\inst{2,3,5}
          }

   \institute{DARK, Niels Bohr Institute, University of Copenhagen, Jagtvej 155A, DK-2200 Copenhagen, Denmark 
        \and
        Cosmic Dawn Center (DAWN), Denmark
        \and 
        DTU Space, Technical University of Denmark, Elektrovej 327, DK-
2800 Kgs. Lyngby, Denmark
        \and 
        Niels Bohr Institute, University of Copenhagen, Jagtvej 128, DK-2200, Copenhagen, Denmark
        \and
        INAF-Osservatorio Astrofisico di Arcetri, Largo Enrico Fermi 5, I-50125 Firenze, Italy
        }


 
  \abstract
   {
   We present \src, a sub-millimetre galaxy (SMG) spectroscopically confirmed at $z=4.051$ with JWST/NIRSpec. Modelling 31 bands of photometry from the rest-frame ultraviolet to the millimetre, we find a stellar mass of $\log(M_\star/\Msun)=11.34^{+0.15}_{-0.22}$, making \src\ one of the most massive spectroscopically confirmed galaxies at $z>4$, yet consistent with $\Lambda$CDM for a star-formation efficiency of $\approx$20\%. 
   Although SMGs are often thought to be building the dense cores of compact quiescent galaxies, \src\ is extended with a rest-frame 0.9\,$\mu$m half-light radius of 4.3\,kpc. Both spatially resolved SED fitting and rest-frame 1.5\,$\mu$m imaging show that its stellar mass is also extended, with a half-mass radius of $3.1\pm0.4$\,kpc. 
   This is three times larger than quiescent galaxies of the same mass, but consistent with the size--mass relation of star-forming galaxies at $4<z<5$ extrapolated to $\rm M_\star=10^{11.3}$\,\Msun. 
   \src\ forms stars on the main sequence ($\approx$400\,\Msun\,yr$^{-1}$) with dust spread across its disc. 
   Both its stellar mass and dust are more extended than those of $z\approx3$ SMGs. Alongside GN20.2b, which has a similar mass and star-formation rate but is three times smaller, these show that massive dusty galaxies at $z\approx4$ can have very different stellar structures, and that not all of them have their stellar mass concentrated in a compact core.

   }

   \keywords{Galaxies: high-redshift, Galaxies: evolution, Galaxies: formation, Galaxies: star-formation, Galaxies: structure}

   \maketitle
%

\section{Introduction}

Sub-millimetre galaxies (SMGs), the most extreme of the dusty star-forming galaxies (DSFGs), are among the most
massive galaxies at any epoch, with $\rm M_\star\sim10^{11}$\,\Msun\ and star-formation rates of $\sim$200--2000\,\Msun\,yr$^{-1}$ hidden behind large columns of dust \citep[for reviews see][]{Casey2014, Hodge2020}. 
Their number density peaks at $z\sim2.5$ \citep{Danielson2017, Dudzeviciute2020}, and they have long been posited as
the progenitors of massive quiescent galaxies \citep[MQGs;][]{Michalowski2010, Toft2014, Valentino2020}.
JWST has now confirmed large numbers of MQGs at $z>3$ \citep{Carnall2023, Baker2025d} including many spectroscopically above $z=4$ \citep[e.g.][]{Carnall2023Nature, Carnall2024, deGraaff2025, Baker2025a, Ito2025b}, and some with $M_\star\geq10^{11}$\,\Msun\ and early formation times \citep{Glazebrook2024, Carnall2024,deGraaff2025}, implying comparably massive progenitors at even earlier times.

Such extremely massive galaxies are powerful tests of galaxy formation, and potentially of cosmology as in
$\Lambda$CDM a galaxy's stellar mass is limited by the mass of its halo, the cosmic baryon fraction and the
efficiency with which those baryons are turned into stars, $M_\star=\epsilon f_{\rm b} M_{\rm h}$, so a given
volume sets a maximum stellar mass \citep{Behroozi2018, Boylan-Kolchin2023}. 
JWST has uncovered
several apparently over-massive galaxies \citep[e.g.][]{Labbe2023, Xiao2024,Xiao2026b, Glazebrook2024, Carnall2024,
deGraaff2025}, but their masses are uncertain, candidates suffer from interlopers, inaccurate redshifts and stellar-population and dust assumptions, whilst the tension for MQGs rests on their inferred star-formation histories which can depend on stellar population models.

The structure of these galaxies can be as informative as their mass. MQGs at $3<z<5$ are remarkably compact, with
half-light radii of only $\approx$0.2--1\,kpc at $M_\star\approx10^{10.5}$--$10^{11}$\,\Msun\
\citep{Ito2024, Carnall2024, Baker2025a}, so their progenitors must have built dense stellar cores rapidly.
The compact ($R_{\rm e}\approx1$--2\, kpc) dust continuum of SMGs has been taken as evidence that they are doing exactly this \citep{Toft2014, Hodge2016, Gullberg2019, Tan2024}. However, the dust continuum mainly traces ongoing star-formation rather than the existing stellar mass, and hence, prior to JWST the mass distribution of the most massive DSFGs at $z>4$ was hidden. 

JWST now resolves their rest-frame
near-infrared light, 
confirming, as suggested by HST at $z\approx2$ \citep{Lang2019}, that their stellar discs are often much larger than the dust \citep{ CrespoGomez2024, Gillman2024}. However, radial gradients in age and attenuation mean that light-weighted sizes can differ significantly from mass-weighted ones
\citep[e.g.][]{Suess2019}. 
Resolved SED fitting of $z\approx3$ SMGs indeed finds stellar masses as compact as their dust \citep[$R_{\rm e,\star}\approx2$\,kpc;][]{Li2026}, but it is unknown where the stellar mass lies in the most massive SMGs at $z>4$.
Whether these galaxies are on the way to becoming compact MQGs therefore requires measuring where their stellar mass (not just their light) is located.

What is lacking are spectroscopically confirmed, $z>4$, $\rm M_\star>10^{11}$\,\Msun\ DSFGs, of which there are
only a handful, including GN20 and GN20.2b \citep{Daddi2009}, and GN10 \citep{Riechers2020}. Here we present and spectroscopically confirm \src, a 
\Msrc\ SMG at $z=4.051$. It was first identified in the sub-millimetre as AzUDS57 \citep{Michalowski2012a}
and localised with ALMA as AS2UDS0036.0 \citep{Stach2019}, with photometric redshifts ranging from
$z=3.4^{+1.1}_{-0.3}$ \citep{Stach2019,Dudzeviciute2020} to $z=4.5^{+0.18}_{-0.15}$ \citep{Ikarashi2017}. 
We measure its stellar mass from 31 bands of photometry and, using spatially resolved SED fitting and MIRI imaging, where that mass is located.
We use a \citet{Chabrier2003} initial mass function (IMF) and \citet{PlanckCollaboration2020} cosmology throughout.

\begin{figure*}
    \centering
    \includegraphics[width=0.99\linewidth]{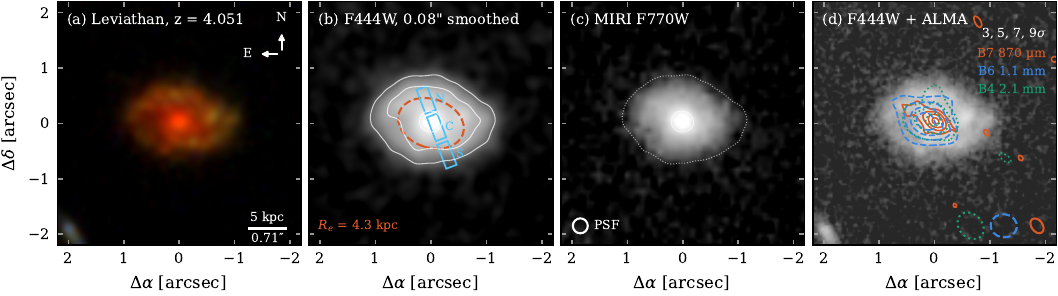}
       \caption{\src\ in JWST and ALMA imaging. (a) NIRCam RGB image (F356W–F480M, F200W–F335M, F090W–F182M) with a 5\,kpc scale bar. (b) F444W, smoothed by $0\farcs08$ to show the low-surface-brightness emission, with the single-S\'ersic half-light ellipse ($R_{\rm e}=4.3$\,kpc), the three NIRSpec shutters (N, C, S) and the PSF. (c) MIRI F770W (rest-frame 1.5\,$\mu$m) with the 3$\sigma$ (dotted) and 30$\sigma$ (solid) contours of the F444W low-surface-brightness stack.
   (d) F444W with ALMA Band 7, 6 and 4 continuum contours (3, 5, 7 and 9$\sigma$), showing that the dust, although extended, is
   more compact than the stellar light.}
    \label{fig:rgb}
\end{figure*}

\section{Methods and Results}

\begin{figure*}
    \centering
    \includegraphics[width=0.99\linewidth]{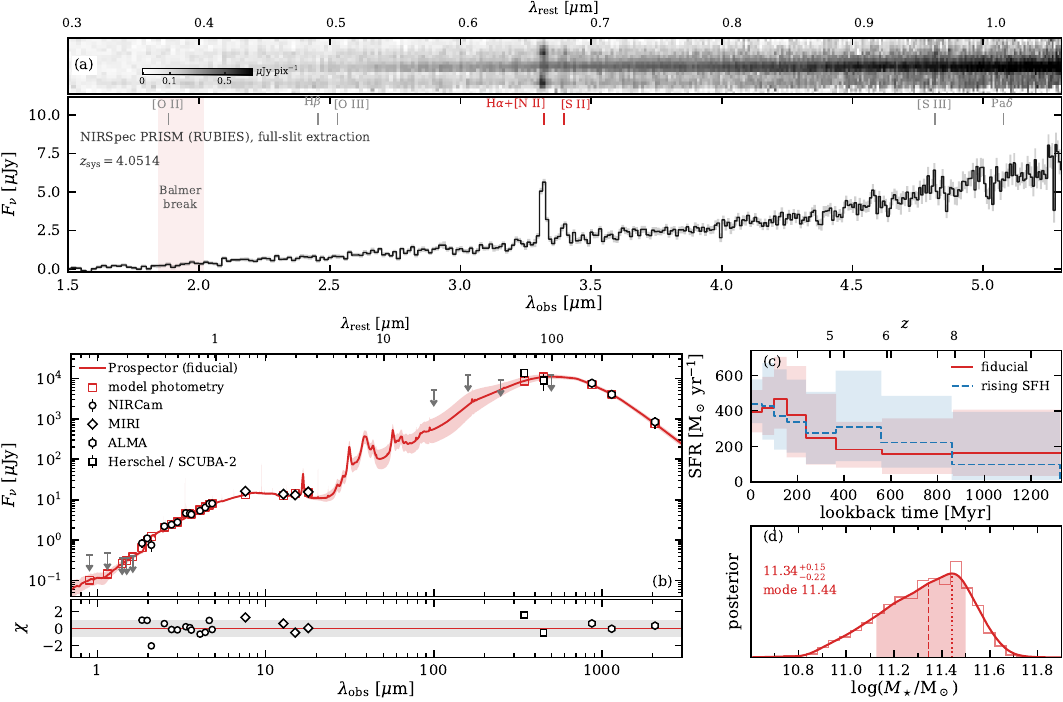}
    \caption{(a) 2D and 1D NIRSpec prism spectrum of \src, with H$\alpha$+[NII] and [SII] confirming the redshift. (b) \prospector\ fit to the 31 bands, with the $\chi$ residuals. (c) Star-formation histories of the two main runs. (d) Stellar-mass posterior.} 
    \label{fig:spec}
\end{figure*}

We use NIRSpec prism and G395M spectroscopy from RUBIES \citep{deGraaff2024}, NIRCam and MIRI imaging from
PRIMER \citep{Donnan2024} and MINERVA \citep{Muzzin2025}, all reduced by the DJA \citep{Brammer2023,
Valentino2023, Heintz2025, Valentino2025}, and far-infrared to millimetre photometry from \textit{Herschel} \citep{Dudzeviciute2020},
SCUBA-2 and ALMA (Appendices \ref{app.data.phot} and \ref{app.data.spec}). \src\ is very red with an average extinction of $A_V\approx2.5$ and $\rm F200W-F444W \approx 1.9 \,mag$.

From H$\alpha$ and [NII] in the G395M spectrum we measure a systemic redshift of $z_{\rm sys}=4.0514\pm0.0017$
(Appendix \ref{app.grating}), consistent with the prism, which shows H$\alpha$+[NII] and [SII] with a strongly
attenuated continuum ($A_V\approx3$) and clear spatial gradients in H$\alpha$ along the slit (Fig. \ref{fig:spec}).

We model all 31 photometric bands, including the far-infrared upper limits, with \prospector\
\citep{Johnson2021}, following \citet{Baker2025a, Baker2026} (Appendix \ref{app.SED_fit}). As the slit covers
only 16\% of the F444W light of \src, we use the prism only to set the redshift prior, since including it would
bias the fit, and we find no evidence for a significant AGN (Appendix \ref{app.agn}). The fit reproduces almost all bands over 1--1100\,$\mu$m (Fig. \ref{fig:spec})
and gives $\log(M_\star/\Msun)=11.34^{+0.15}_{-0.22}$. A rising-SFH prior, which can lower masses
\citep{Lapasia2026}, gives a consistent $\rm log(M_\star/M_\odot)=11.23^{+0.16}_{-0.20}$. The SFR of
$410^{+130}_{-90}$\,\Msun\,yr$^{-1}$ places \src\ on the massive end of the star-forming main sequence at $z\approx4$ \citep{Popesso2023}. 
Fitting the far-infrared SED alone with \textsc{mercurius} \citep{Witstok2022} (Appendix~\ref{app.fir}) gives $T_{\rm dust}\approx30$\,K and $M_{\rm dust}=2.6^{+1.4}_{-0.8}\times10^9$\,\Msun, a dust-to-stellar mass ratio of 1.2\%.

\begin{table}
\caption{Properties of \src.}
\label{tab:props}
\centering
\small
\begin{tabular}{lc}
\hline\hline
Quantity & Value \\
\hline
RA, Dec (J2000) & 02:17:27.31, $-$05:06:42.9 \\
$z_\mathrm{sys}$ (G395M H$\alpha$); $z_\mathrm{prism}$ & $4.0514\pm0.0017$; $4.049\pm0.004$ \\
$\log(M_\star/\mathrm{M}_\odot)$ & $11.34^{+0.15}_{-0.22}$ (mode 11.44)$^a$ \\
SFR$_{100}$ [M$_\odot$\,yr$^{-1}$] & $410^{+130}_{-90}$ \\
$t_{50}$ (lookback) [Myr] & $435^{+317}_{-225}$ \\
$A_V$ [mag] & $2.5^{+0.2}_{-0.3}$ \\
$L_\mathrm{IR}$ [$10^{12}$\,L$_\odot$]$^b$ & $5.6^{+2.1}_{-1.4}$ \\
$T_\mathrm{dust}$ [K]$^c$ & $29^{+3}_{-3}$ \\
$M_\mathrm{dust}$ [$10^9$\,M$_\odot$]$^c$ & $2.6^{+1.4}_{-0.8}$ \\
$F(\mathrm{H}\alpha+[\mathrm{N\,II}])$ [$10^{-17}$\,cgs]; EW$_0$ [\AA]$^d$ & $0.79\pm0.03$; $119$ \\
$R_\mathrm{e}$ (F444W, major axis) [kpc]$^e$ & $4.3\pm0.2$ \\
S\'ersic $n$; axis ratio $q$ & $1.18\pm0.02$; $0.72$ \\
$R_\mathrm{e}$ (F770W, major axis) [kpc]$^f$ & $3.3\pm0.3$ \\
$R_\mathrm{e,mass}$ (resolved SED fits) [kpc]$^g$ & $3.1\pm0.4$ \\
$R_\mathrm{e,dust}$ (ALMA B6) [kpc]$^h$ & $2.2\pm0.2$ \\
$R_\mathrm{e,dust}$ (ALMA B4) [kpc]$^h$ & $2.5\pm0.2$ \\
\hline
\end{tabular}
\tablefoot{$M_\star$ is the surviving mass.
\tablefoottext{a}{The posterior has a low-mass tail, so the median lies below the mode.}
\tablefoottext{b}{Energy balance, 8--1000\,$\mu$m.}
\tablefoottext{c}{Optically thin greybody (\textsc{mercurius}); $\lambda_0=100\,\mu$m gives $34\pm4$\,K and
$2.0^{+1.0}_{-0.6}\times10^9$\,M$_\odot$ (App.~\ref{app.fir}).}
\tablefoottext{d}{NIRSpec prism, in slit}
\tablefoottext{e}{Range of PSF choices and fitting methods (App.~\ref{app.morph}).}
\tablefoottext{f}{Rest-frame 1.5\,$\mu$m; single S\'ersic fit, $n\approx1.0$ (App.~\ref{app.morph}).}
\tablefoottext{g}{Half-mass radius.}
\tablefoottext{h}{Gaussian fits in the uv plane; Band 6 from \citet{Ikarashi2017} (App.~\ref{app.morph}).}}
\end{table}

Fitting the light profile with a single S\'ersic model in \textsc{pysersic} \citep{Pasha2023} gives a half-light radius of
$4.3\pm0.2$\,kpc in F444W \citep[consistent with][]{Gillman2024}, decreasing to $3.3\pm0.3$\,kpc in F770W (rest-frame
1.5\,$\mu$m; Appendix \ref{app.morph}). To measure where the stellar mass is, we bin PSF-matched NIRCam images
with \textsc{piXedfit} \citep{Abdurrouf2021} and fit each region with \prospector\ (Appendix \ref{app.outshining}). 
This gives a half-mass radius of $3.1\pm0.4$\,kpc, in agreement with the independent half-light radius at rest-frame 1.5\,$\mu$m ($3.3\pm0.3$\,kpc in F770W), where the light is less affected by dust and is a much closer tracer of stellar mass. The dust  continuum has
$R_{\rm e}\approx2.2$--2.5\,kpc (uv-plane fits in Bands 6 and 4) making it more compact than the stars but larger than the typical size of SMGs \citep[$\approx$1--2\,kpc,][]{Hodge2016, Gullberg2019}. The summed mass of the regions agrees with our fiducial value, whereas a NIRCam-only fit to the integrated photometry gives a mass 0.3 dex lower, consistent with outshining by the less obscured outskirts \citep{Michalowski2012}. Our
fiducial fit recovers the resolved mass  (Table~\ref{tab:appB}).

\section{Discussion}

\begin{figure*}
    \centering
    \includegraphics[width=0.99\linewidth]{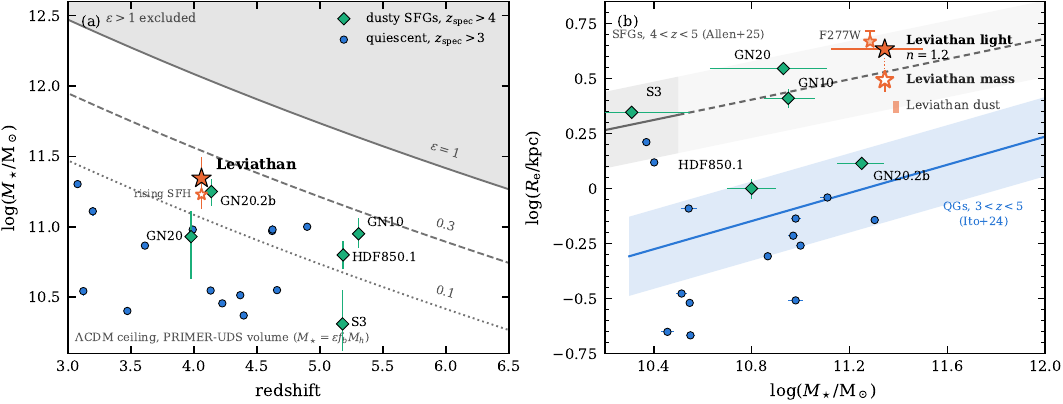}
    \caption{\src\ in context. (a) Stellar mass against redshift. The lines show the maximum stellar mass
    expected in the PRIMER-UDS volume for $M_\star=\epsilon f_{\rm b}M_{\rm h}$ with $\epsilon=1$, 0.3 and 0.1,
    with $\epsilon>1$ shaded. \src\ is shown for our fiducial fit (filled star) and with the rising-SFH prior
    (open star). We compare to spectroscopically confirmed dusty star-forming galaxies at $z>4$ (diamonds; GN10 and
    S3, \citealt{Lapasia2026}; HDF850.1, \citealt{Sun2024}; GN20 and GN20.2b, \citealt{CrespoGomez2024}) and
    massive quiescent galaxies at $z>3$ (circles; \citealt{Carnall2023, Carnall2024, deGraaff2025, Baker2025a}). GN20 and
    GN20.2b are offset by $\pm0.08$ in $z$ for clarity. (b) Size against stellar
    mass. \src\ is shown for its F444W half-light radius (filled star), its half-mass radius from the resolved
    SED fits (open star; Appendix \ref{app.outshining}) and its ALMA dust size (bar), with the dusty galaxies
    and quiescent galaxies of panel (a), the size--mass relation of quiescent
    galaxies at $3<z<5$ \citep{Ito2024}, and that of star-forming galaxies at $4<z<5$ \citep[grey;][dashed
    where extrapolated beyond $\log(M_\star/\Msun)\approx10.5$]{Allen2025}.}
    \label{fig:context}
\end{figure*}

Fig. \ref{fig:context}a) compares \src\ to spectroscopically confirmed DSFGs and MQGs at $z>3$, with the
maximum stellar mass expected in the PRIMER-UDS volume \citep{Behroozi2018, Boylan-Kolchin2023}. 
\src\ is one of the most massive
spectroscopically confirmed galaxies at $z>4$, but would require $\epsilon\approx20$\% (10--30\% within its
mass uncertainty) if it occupies this halo, posing no tension with $\Lambda$CDM.

The mass-size relation is currently only reliably measured up to $M_\star\approx10^{10.5}$\,\Msun at $z\sim4$ \citep[e.g.][]{Allen2025}.
In the size--mass plane (Fig. \ref{fig:context}b), \src\ lies within 1$\sigma$ of the size--mass relation
of star-forming galaxies at $4<z<5$ \citep{Allen2025} extrapolated to its mass, both in light (4.6\,kpc in
F277W, at the same rest-frame wavelength and also measured along the major axis) and in stellar mass. 
As photometric masses
are uncertain at the massive end (Appendix \ref{app.outshining}), \src\ extends the comparison to above $10^{11}$\,\Msun\ suggesting that the relation may not flatten up to $10^{11.3}M_\odot$. It is also larger than most SMGs at $z\approx2.7$
\citep[$R_{\rm e}=2.7^{+1.7}_{-0.9}$\,kpc at $M_\star\approx10^{11.2}$\,\Msun;][]{Gillman2024} and its half-mass radius is 1.6 times that of $z\approx3$ SMGs from resolved SED fitting \citep{Li2026}. 
GN20.2b, at the same redshift with a similarly high stellar mass and SFR, is three
times smaller \citep{CrespoGomez2024}, so massive dusty galaxies at $z\approx4$ can have very different stellar structures.

The differences between the half-light, half-mass and dust radii of \src\ can help constrain how it is growing. Its half-mass radius is
$0.72\pm0.08$ times its half-light radius, typical of inside-out growing massive star-forming galaxies at cosmic noon, in which the
centre is older and dustier than the outskirts \citep{Suess2019}, as the prism spectra confirm for \src\
(Appendix \ref{app.data.spec}). 
The dust is only slightly more compact than the stellar mass ($R_{\rm e,dust}\approx2.2$--2.5\,kpc). Temperature gradients can make the continuum appear more concentrated than the dust mass \citep{Boogaard2026}, so the star formation may be even more extended.
The star formation is  spread across the disc
with $\Sigma_{\rm SFR}\approx10$\,\Msun\,yr$^{-1}$\,kpc$^{-2}$, compared to $\rm \approx65\,\Msun\,yr^{-1}\,kpc^{-2}$ for GN20 at a similar dust
size. Even if the gas reservoir of \src\ ($M_{\rm gas}\sim1$--$3\times10^{11}$\,\Msun\ for gas-to-dust ratios of 50--100)
were converted into stars with the present distribution of the dust, adding 0.3\,dex in mass, the half-mass
radius would only decrease to $\approx2.8$\,kpc. The central density implied by our profile,
$\log(\Sigma_{\rm1,kpc}/\Msun\,{\rm kpc}^{-2})\approx9.9$, is also below that of compact quiescent galaxies, although a
compact component below our $\approx$1.5\,kpc PSF matched resolution limit cannot be excluded. If its star formation continues
with its present spatial distribution, \src\ is therefore unlikely to develop the compact structure of the
massive quiescent galaxies at $3<z<5$; this would require substantial central mass growth or a structural
transformation, for example through gas inflows or mergers. 
Whereas the stellar mass of $z\approx3$ SMGs is typically compact ($R_{\rm e,\star}\approx2$\,kpc; \citealt{Li2026}), that of \src\ is extended at a higher mass, as is its dust.
Together with GN20.2b, this suggests that massive galaxies at $z\approx4$ follow different structural paths, with \src\ a natural progenitor of the extended, disc-like quiescent galaxies at $z\approx2$--3 \citep[e.g.][]{Toft2017}.
\src\ lies in a modest, mass-biased overdensity with a $\log(M_\star/\Msun)\approx10.3$ companion at 23\,kpc (Appendix~\ref{app.env}). Whether it becomes the central galaxy of a group or a brightest cluster galaxy is an open question.

\section{Conclusions}

\begin{itemize}
    \item We spectroscopically confirm \src\ at $z=4.051$ with $\log(M_\star/\Msun)=11.34^{+0.15}_{-0.22}$, robust to within 0.2\,dex, making it one of the most massive spectroscopically confirmed galaxies at $z>4$, yet consistent with $\Lambda$CDM for $\epsilon\approx20$\%.
    \item Both its light ($R_{\rm e}=4.3$\,kpc) and stellar mass ($R_{\rm e,mass}=3.1$\,kpc) are extended, consistent with the extrapolated size–mass relation of star-forming galaxies and three times larger than quiescent galaxies of the same mass, with the dust only slightly more compact.
    \item With its present distribution of star formation, \src\ is unlikely to become a compact quiescent galaxy. With GN20.2b, it shows that the most massive dusty galaxies at $z\approx4$ can have very different stellar structures.
\end{itemize}

\bibliographystyle{aa}
\bibliography{refs}

\begin{appendix}
 
\section{Photometry}
\label{app.data.phot}
\begin{table}
\caption{Photometry of \src\ used in the SED fits. NIRCam values are totals in an elliptical aperture
($a=2\farcs45$, $b=1\farcs96$); bands below $2\sigma$ are given as $2\sigma$ limits (they are fitted as
measurements). Errors are as measured; the fits use $\max(\sigma, 0.05F_\nu)$.
Far-IR limits are $3\sigma$ and enter the fits as censored constraints.}
\label{tab:phot}
\centering
\small
\setlength{\tabcolsep}{4pt}
\begin{tabular}{lc|lc}
\hline\hline
\multicolumn{2}{l|}{\textit{NIRCam} [$\mu$Jy]} & & \\
\hline
F090W & $<0.44$ & \multicolumn{2}{l}{\textit{MIRI} [$\mu$Jy]} \\
F115W & $<0.47$ & F770W & $16.2\pm2.2$ \\
F140M & $<0.35$ & F1280W & $13.7\pm1.8$ \\
F150W & $<0.36$ & F1500W & $13.1\pm2.1$ \\
F162M & $<0.40$ & F1800W & $15.5\pm4.2$ \\
F182M & $0.84\pm0.20$ & \multicolumn{2}{l}{\textit{far-IR / mm} [mJy]} \\
F200W & $1.11\pm0.14$ & PACS 100\,$\mu$m$^a$ & $<5.2$ \\
F210M & $0.75\pm0.24$ & PACS 160\,$\mu$m$^a$ & $<12.1$ \\
F250M & $2.20\pm0.20$ & SPIRE 250\,$\mu$m$^a$ & $<9.2$ \\
F277W & $2.41\pm0.17$ & SPIRE 350\,$\mu$m$^a$ & $13.7\pm3.2$ \\
F300M & $2.77\pm0.29$ & SCUBA-2 450\,$\mu$m$^b$ & $8.9\pm4.0$ \\
F335M & $4.66\pm0.24$ & SPIRE 500\,$\mu$m$^a$ & $<12.3$ \\
F356W & $4.62\pm0.26$ & ALMA B7, 344\,GHz$^c$ & $7.60\pm1.10$ \\
F360M & $4.31\pm0.28$ & ALMA B6, 262\,GHz$^c$ & $4.02\pm0.63$ \\
F410M & $5.32\pm0.32$ & ALMA B4, 145\,GHz$^c$ & $0.84\pm0.27$ \\
F444W & $6.40\pm0.37$ & \multicolumn{2}{l}{\textit{not fitted}$^d$} \\
F460M & $7.98\pm0.61$ & SCUBA-2 850\,$\mu$m & $7.5\pm1.2$ \\
F480M & $8.05\pm0.42$ & SMA 860\,$\mu$m & $6.6^{+1.1}_{-1.0}$ \\
 &  & ALMA B4 (2023) & $0.70\pm0.15$ \\
\hline
\end{tabular}
\tablefoot{
\tablefoottext{a}{\citet{Dudzeviciute2020}, AS2UDS\,0036.0.}
\tablefoottext{b}{This work: 390 JCMT archive nightly maps, matched filter; error scaled by the null test (App.~\ref{app.fir}).}
\tablefoottext{c}{Model fits to the images (curve of growth for B7); B6 is the weighted mean of 2013.1.00781.S and 2015.1.00442.S; B6 and B4 errors include 10\% calibration. Programmes: B7 2015.1.01528.S, 2016.1.00434.S; B4 2021.1.00705.S.}
\tablefoottext{d}{Consistency checks: S2CLS and SMA (UDS40, deboosted; \citealt{Hill2018}); ALMA 2023.1.01520.S (this work).}}
\end{table}

\subsection{NIRCam and MIRI}
We use the DJA v7.2 mosaics \citep{Brammer2023,Valentino2023} of PRIMER-UDS \citep{Donnan2024} (Dunlop+, in prep) and MINERVA \citep{Muzzin2025}. We detect sources on a multi-band detection image, and replace the segments of neighbouring sources (dilated by the PSF FWHM) with the mirror image of \src\ through its centroid. 
We keep the bright knot 0\farcs59 to the west as part of \src, as the F444W emission between the two does not fall to the sky level and the knot lies within the ALMA emission. 
We measure total fluxes in an
elliptical aperture whose shape is set by the second moments of the detection image and whose size is set by the curve of growth ($a=2\farcs45$, $b=1\farcs96$). 
For an aperture this large the colour error from the different PSFs is only 0.05\%, so we do not PSF-match the images (except in the spatially resolved analysis). The MIRI fluxes are measured at the native resolution, with aperture corrections from forward-modelling the source through the MIRI PSFs. 
The uncertainties come from empty apertures, and in the SED fitting we adopt an error floor of 5\% in all bands \citep{Baker2025a}. Our F444W flux ($6.40\pm0.37\,\mu$Jy) agrees with the DJA catalogue (6.3\,$\mu$Jy), whilst our photometry is bluer at $\lambda<2\,\mu$m, as expected for a red core surrounded by bluer outskirts.
 
\subsection{FIR and sub-mm}
We take the \textit{Herschel} PACS and SPIRE fluxes from \citet{Dudzeviciute2020}. \src\ is only detected
at 350\,$\mu$m ($13.7\pm3.2$\,mJy), and we include the $3\sigma$ upper limits in the other bands as
censored constraints in all of our fits. As \textit{Herschel} only weakly constrains the peak of the FIR
SED, we measure the SCUBA-2 450\,$\mu$m flux from the 390 nightly maps in the JCMT Science Archive that
cover \src, using a point-source matched filter on each map and
combining them with inverse-variance weights. 
This gives $8.86\pm1.64$\,mJy, and we inflate the
uncertainty to $\pm4.0$\,mJy based on the scatter of the same measurement at 20 empty positions.
 
As aperture photometry on interferometric images is biased by negative bowls, we fit the ALMA Band 6 and 4 images with an elliptical Gaussian convolved with the restoring beam, adding a 10\% calibration uncertainty in quadrature\footnote{\href{https://doi.org/10.5281/zenodo.18793803}{ALMA Technical Handbook}}. 
Combining the two Band 6 datasets gives $4.02\pm0.63$\,mJy, and we find $0.84\pm0.27$\,mJy in Band 4. In Band 7 the curve of growth reaches a stable plateau beyond $\sim$1\arcsec, giving $7.60\pm1.10$\,mJy, significantly higher than the catalogue value of \citet{Stach2019} (4.73\,mJy), which does not recover the extended emission of \src. 
The deboosted SCUBA-2 850\,$\mu$m and SMA 860\,$\mu$m fluxes \citep{Hill2018} and a more recent Band 4 map ($0.70\pm0.15$\,mJy), which we do not fit, all agree with our best-fit FIR model (Appendix \ref{app.fir}).
 
\section{NIRSpec spectroscopy}
\label{app.data.spec}
 
\subsection{Prism}
The RUBIES \citep{deGraaff2024} slitlet (PA $=200\fdg7$) lies close to the minor axis of \src, crossing the core and clumps to
the north and south (Fig. \ref{fig:rgb}). As the galaxy fills the slit, we re-extract the prism spectrum
from the DJA two-dimensional products \citep{Heintz2025} by summing all of the rows covered by \src, with a global sky
subtraction. We fit the emission lines with Gaussians at the prism resolution (scaled by a single factor
fitted on \ha) and a quadratic continuum, fixing [NII]$\lambda6584$/\ha$=0.45$ from G395M, with
uncertainties from 300 Monte Carlo realisations. We detect \ha+[NII] (rest-frame EW of 119\,\AA), [SII] and
marginally [SIII]$\lambda9531$, whilst H$\beta$, [OIII] and [OII] are not detected, as expected given the
heavy attenuation. The prism gives $z=4.049\pm0.004$. Comparing synthetic photometry of the spectrum with
PSF-matched photometry inside the three shutters gives a grey scale factor of $3.67\pm0.06$, and hence an
in-slit \ha+[NII] flux of $(7.9\pm0.4)\times10^{-18}$\,erg\,s$^{-1}$\,cm$^{-2}$. The shutters contain only 16\%
of the F444W flux, so we use the prism only for the redshift. Fitting the spectrum
jointly with the photometry can only reconcile the two by driving the metallicity to $\sim$0.1\,Z$_\odot$.
 
\subsection{G395M}
\label{app.grating}
\begin{table}
\caption{H$\alpha$ and [N\,II]$\lambda6585$ in the three G395M shutters (fluxes in $10^{-18}$\,erg\,s$^{-1}$\,cm$^{-2}$,
inside the shutter). $\Delta v$ is relative to $z=4.0514$.}
\label{tab:g395m}
\centering
\small
\setlength{\tabcolsep}{4pt}
\begin{tabular}{lcccc}
\hline\hline
Shutter & $\Delta v$ [km\,s$^{-1}$] & H$\alpha$ & [N\,II] & [N\,II]/H$\alpha$ \\
\hline
N$^a$ & $-337\pm18$ & $2.10\pm0.21$ & $1.18\pm0.37$ & $0.56^{+0.14}_{-0.08}$ \\
C & $-2\pm41$ & $1.00\pm0.19$ & $0.54\pm0.37$ & $0.54^{+0.24}_{-0.15}$ \\
S & $+10\pm28$ & $1.72\pm0.22$ & $0.43\pm0.32$ & $0.25^{+0.20}_{-0.06}$ \\
\hline
Sum$^b$ & -- & $4.83\pm0.35$ & $2.15\pm0.60$ & $0.44$ \\
\hline
\end{tabular}
\tablefoot{Statistical errors; the grey photometric scale adds 45\%, 10\% and 26\% (S, C, N).
\tablefoottext{a}{Contains clump A.}
\tablefoottext{b}{H$\alpha$+[N\,II] $=7.70\pm0.92$, compared to
$7.90\pm0.35$ from the full-slit prism divided by 3.67.}}
\end{table}

For G395M we use spectra extracted separately for the south, centre and north shutters with the master background (as the galaxy fills the slitlet, nod subtraction would remove part of it). We then subtract the residual background measured in the outermost, source-free nodded shutter position, which removes a residual sky offset of $\approx$1\,$\mu$Jy at 2.9--3.6\,$\mu$m, and scale each spectrum to the PSF-matched F410M and F444W photometry in the same footprint.
We fit \ha\ and [NII]$\lambda\lambda6549,6585$
with a common redshift and width at the G395M resolution, limiting the intrinsic dispersion to
$\leq300$\,km\,s$^{-1}$, with uncertainties from 200 Monte Carlo realisations (Table \ref{tab:g395m}).  
The line widths are unresolved in all three shutters ($\sigma_{\rm int}\lesssim110$\,km\,s$^{-1}$, below the G395M resolution).
The central and southern shutters give $z=4.0514\pm0.0007$
and $4.0516\pm0.0005$, and we adopt $z_{\rm sys}=4.0514\pm0.0017$, including a $\sim$100\,km\,s$^{-1}$
zero-point uncertainty for extended emission. The northern clump is blueshifted by
$337\pm18$\,km\,s$^{-1}$, but with a single slit close to the minor axis we do not interpret this as rotation.
The summed \ha+[NII] flux, $(7.7\pm0.9)\times10^{-18}$\,erg\,s$^{-1}$\,cm$^{-2}$, agrees with the in-slit
prism value.
 
\section{FIR SED}
\label{app.fir}
\begin{figure}
    \centering
    \includegraphics[width=0.99\linewidth]{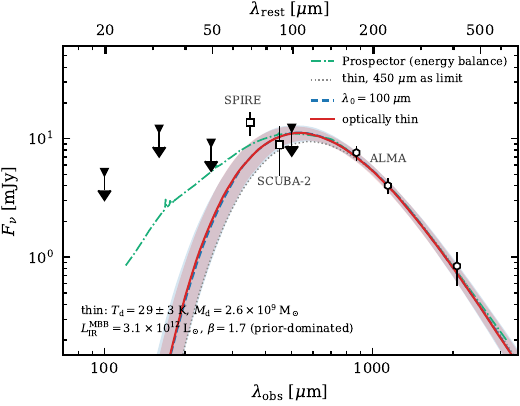}
    \caption{FIR to mm SED of \src. Open symbols show the SPIRE 350\,$\mu$m, SCUBA-2 450\,$\mu$m and
    ALMA detections, and triangles the $3\sigma$ upper limits from PACS and SPIRE. The red solid and blue
    dashed lines (with 16th--84th percentile ranges) show the optically thin and general opacity
    ($\lambda_0=100\,\mu$m) modified-blackbody fits, and the grey dotted line the optically thin fit with
    450\,$\mu$m as an upper limit. The green dot-dashed line shows the fiducial \prospector\ SED.}
    \label{fig:fir}
\end{figure}
 
We fit the FIR and sub-mm photometry with modified blackbodies using \textsc{Mercurius}
\citep{Witstok2022}, with the same detections and censored limits as our \prospector\ fits,  with detections covering rest-frame 70--410\,$\mu$m. We include the effects of the CMB and use the default
\textsc{Mercurius} opacity and priors (including $\beta=1.8\pm0.25$). For an optically thin model we find
$T_{\rm dust}=29.0^{+3.3}_{-3.1}$\,K, $M_{\rm dust}=2.6^{+1.4}_{-0.8}\times10^{9}$\,\Msun\ and
$L_{\rm IR}=(3.1\pm0.7)\times10^{12}$\,L$_\odot$, whilst a general opacity model ($\lambda_0=100\,\mu$m)
gives $T_{\rm dust}=34\pm4$\,K and $M_{\rm dust}=2.0^{+1.0}_{-0.6}\times10^{9}$\,\Msun\ with the same $L_{\rm IR}$.
The emissivity index is set by the prior. The luminosity corresponds to an obscured SFR of
$\sim$470\,\Msun\,yr$^{-1}$, in good agreement with \prospector. Treating
450\,$\mu$m as an upper limit or removing it changes the dust mass by $\leq0.15$\,dex, but the temperature
is only constrained to $\sim$25--35\,K. The \prospector\ $L_{\rm IR}$ ($5.6\times10^{12}$\,L$_\odot$) is higher
as it also includes warm dust and PAH emission at $\lambda_{\rm rest}\lesssim50\,\mu$m, which a
single-temperature greybody does not capture.
 
\section{Prospector fitting}
\label{app.SED_fit}
\begin{table}
\caption{Robustness of the stellar mass to the fitting choices (medians; $M_\star$ with 16th--84th
percentiles). All runs are photometry-only with a continuity SFH unless noted.}
\label{tab:appB}
\centering
\small
\setlength{\tabcolsep}{3pt}
\begin{tabular}{lcccc}
\hline\hline
Run & $\log(M_\star/\mathrm{M}_\odot)$ & SFR$_{100}$ & $t_{50}$ & $L_\mathrm{IR}$ \\
 & & [M$_\odot$\,yr$^{-1}$] & [Myr] & [$10^{12}$\,L$_\odot$] \\
\hline
Fiducial & $11.34^{+0.15}_{-0.22}$ & 408 & 435 & 5.6 \\
450\,$\mu$m as limit & $11.34^{+0.15}_{-0.16}$ & 428 & 534 & 5.9 \\
No 450\,$\mu$m & $11.25^{+0.13}_{-0.29}$ & 446 & 315 & 6.0 \\
Rising SFH & $11.23^{+0.16}_{-0.20}$ & 513 & 236 & 6.3 \\
$Z_\star=\mathrm{Z}_\odot$ & $11.15^{+0.14}_{-0.14}$ & 475 & 253 & 4.5 \\
No MIRI$^a$ & $11.27^{+0.24}_{-0.20}$ & 569 & 297 & 7.1 \\
No far-IR/mm & $11.37^{+0.13}_{-0.25}$ & 245 & 614 & 2.9 \\
Limits as $0\pm\sigma$$^a$ & $11.29^{+0.12}_{-0.19}$ & 321 & 532 & 3.7 \\
\hline
\end{tabular}
\tablefoot{\tablefoottext{a}{With the 450\,$\mu$m limit rather than the measurement.}}
\end{table}

We model the photometry of \src\ with \prospector\ \citep{Johnson2021}, using \textsc{FSPS}
\citep{Conroy2009} with the MILES library, MIST isochrones and a \citet{Chabrier2003} IMF. We use the non-parametric continuity SFH
\citep{Leja2019} with eight bins (0--50 and 50--100\,Myr, then six logarithmic bins to $z=20$) and a
Student-$t$ prior on the log SFR ratios (width 0.3), and report the surviving
stellar mass. The stellar metallicity is free ($-1.5<\log(Z_\star/{\rm Z}_\odot)<0.19$). We use the
\citet{Kriek2013} attenuation law with a free optical depth and slope, plus extra attenuation of stars
younger than 10\,Myr, and \citet{Draine2007} dust emission with $U_{\rm min}$, $q_{\rm PAH}$ and $\gamma$ free,
through energy balance. We include nebular emission with a free ionisation parameter and
gas-phase metallicity, and IGM absorption, but no AGN (Appendix \ref{app.agn}). The
redshift has a Gaussian prior centred on $z=4.058$ ($\sigma=0.02$), and returns
$z=4.056^{+0.014}_{-0.016}$.
 
For the detections we use a Gaussian likelihood with an error floor of 5\%,
$\sigma_i=\max(\sigma_{i,\rm obs},0.05F_i)$ \citep[e.g.][]{Baker2025a}. The undetected \textit{Herschel} bands constrain the Wien side
of the dust emission, but treating them as measurements of zero flux penalises every model with positive
flux and biases the fit towards a fainter FIR SED. Instead, we treat them as censored data where a $3\sigma$ limit $L_j$ contributes $\ln\Phi[(L_j-m_j)/\sigma_j]$ to the
log-likelihood, with $\sigma_j=L_j/3$, $m_j$ the model flux and $\Phi$ the standard normal cumulative distribution. 
This is flat for models well below the limit and strongly penalises models above it. As
shown in Table \ref{tab:appB}, the choice matters for the FIR luminosity (zero-flux limits lower $L_{\rm IR}$
from 5.9 to $3.7\times10^{12}$\,L$_\odot$) but changes the stellar mass by only 0.05\,dex, and removing all of
the FIR and mm data leaves the mass unchanged.
 
Our fiducial fit gives $\log(M_\star/\Msun)=11.34^{+0.15}_{-0.22}$, with the posterior peaking at 11.44 and
only 6\% below $10^{11}$\,\Msun. The low-mass tail is a young, burst-dominated solution, which the SCUBA-2
450\,$\mu$m measurement suppresses (it contains 18\% of the posterior without it). Across all of our
variants the median mass lies between $10^{11.15}$ and $10^{11.37}$\,\Msun\ (Table \ref{tab:appB}). A rising-SFH
prior \citep{Lapasia2026} lowers the mass by 0.11\,dex with an almost identical evidence
($\Delta\ln Z=-0.1$), whilst fixing the metallicity to solar gives the lowest mass (11.15) but is
disfavoured ($\Delta\ln Z=-1.1$). Spatially resolved fits independently give a consistent mass
(Appendix \ref{app.outshining}).
 
\section{Outshining and the half-mass radius}
\label{app.outshining}
\begin{figure}
    \centering
    \includegraphics[width=\columnwidth]{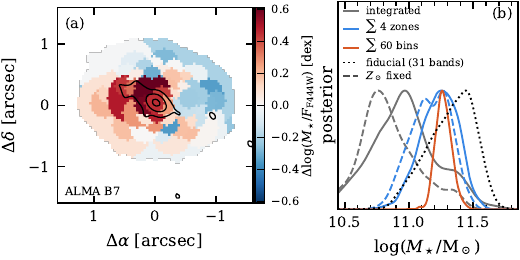}
    \caption{Outshining in \src\ from spatially resolved NIRCam-only \prospector\ fits. (a) Mass per unit
    F444W flux in each of the 60 bins, relative to a NIRCam-only fit to the integrated photometry of the same
    region, with ALMA Band 7 contours. (b) Stellar-mass posteriors for the integrated NIRCam-only fit and the
    sums over four radial zones and 60 bins, with free (solid) and solar (dashed) metallicity, and our
    our fiducial fit (dotted).}
    \label{fig:outshining}
\end{figure}
 
We PSF-match the 18 NIRCam bands to 0\farcs21 and divide \src\ into 60 bins with \textsc{piXedfit}
\citep{Abdurrouf2021}, and into four radial zones. We fit each region with our \prospector\ setup, but with
the rising-SFH prior, the redshift fixed and the dust emission shape fixed, and fit the integrated NIRCam
photometry of the same region identically. The integrated fit gives
$\log(M_\star/\Msun)=10.96^{+0.27}_{-0.24}$, whilst the bins sum to $11.27\pm0.06$. The highest mass-to-light
ratios lie in the central $\sim$4.5\,kpc, coinciding with the dust (Fig. \ref{fig:outshining}), so the
integrated fit is dominated by the less obscured outskirts. The four zones give the same sum (11.24), as
does fixing the metallicity ($+0.37$\,dex), so the offset is not caused by summing many prior-dominated
fits, although with the broad integrated posterior it is only moderately significant (86\% probability).
The resolved mass agrees with our fiducial 31-band fit, which is therefore not significantly biased by outshining.
 
We also measure the half-mass radius, distributing the mass of each bin over its pixels and finding the
ellipse that encloses half of the total, and the half-light radius of the F444W emission in the same way.
The ratio, $R_{\rm e,mass}/R_{\rm e,light}=0.72\pm0.08$, is insensitive to the PSF and is the same for the
four zones (0.72) and with solar metallicity (0.76). Applied to our S\'ersic size this gives
$R_{\rm e,mass}=3.1\pm0.4$\,kpc, which is if anything an underestimate as the bins exclude the faint outskirts.
 
\section{Morphology}
\label{app.morph}
\begin{figure}
    \centering
    \includegraphics[width=\columnwidth]{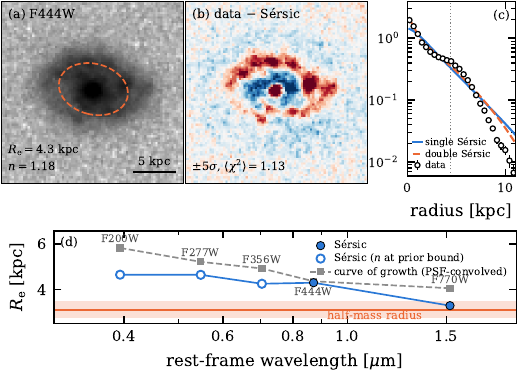}
    \caption{Morphology of \src. (a) F444W image ($3\farcs2$ across) with the half-light ellipse. (b) Residual
    of the single-S\'ersic fit ($\pm5\sigma$), dominated by a clumpy ring at 3--5\,kpc. (c) F444W
    surface-brightness profile (arbitrary units) along the major axis, with the single and double S\'ersic
    models and $R_{\rm e}$ (dotted). (d) Half-light radius against rest-frame wavelength from S\'ersic fits
    (open where $n$ reaches the lower bound of its prior) and from the curve of growth (PSF-convolved),
    compared to the half-mass radius (orange; Appendix \ref{app.outshining}).}
    \label{fig:morph}
\end{figure}
 
We fit F444W with a single Sérsic profile using \textsc{pysersic} \citep{Pasha2023} on a 5$''$ stamp with neighbouring sources masked, using an empirical PSF built from stars in the same mosaic. This gives $R_{\rm e}=4.37$\,kpc, $n=1.18$ and $q=0.72$, and the independent DJA PSFEx PSF\footnote{\url{https://dawn-cph.github.io/dja/blog/2024/08/16/morphological-data/}} gives $R_{\rm e}=4.34$\,kpc and $n=1.19$. As the formal uncertainties of light-profile fits are underestimated, we adopt $R_{\rm e}=4.3\pm0.2$\,kpc (3.7\,kpc circularised), $n=1.2$ and $q=0.72$.
The residuals show a central excess, the clumpy ring and a
steeper-than-exponential decline beyond $\sim$8\,kpc, but a non-parametric curve-of-growth radius on a
16\arcsec\ cutout gives the same size (4.3--4.4\,kpc). A double Sérsic fit gives a compact (0.3 kpc) component with 6\% of the light and a disc with $R_{\rm e}=4.0$\,kpc (Fig. F.1c), so we adopt the single Sérsic size.
 
Repeating the fit in F200W--F356W and MIRI F770W, with empirical PSFs for each image, the size decreases with wavelength, from $\approx$4.6\,kpc at rest-frame 0.55\,$\mu$m to 3.3\,kpc at 1.5\,$\mu$m (Fig. \ref{fig:morph}d), consistent with the half-mass radius. At $\lambda_{\rm rest}\lesssim0.7\,\mu$m the S\'ersic index reaches the lower bound of its prior, so there the curve-of-growth radii (0.5--1.2\,kpc larger) are more reliable. 
The dust is more compact. Gaussian fits in the uv plane give $R_{\rm e,dust}=2.2\pm0.2$\,kpc in Band 6 \citep{Ikarashi2017} and $2.5\pm0.2$\,kpc ($0\farcs35\pm0\farcs03$) in Band 4, consistent with our image-plane fits (FWHM 0\farcs63 and 0\farcs83 for the two Band 6 datasets), which both prefer
an extended component over a compact
(0\farcs15) core, which contains $<$11--22\% of the flux (95\%).
 
\section{Environment}
\label{app.env}
\begin{figure}
    \centering
    \includegraphics[width=\columnwidth]{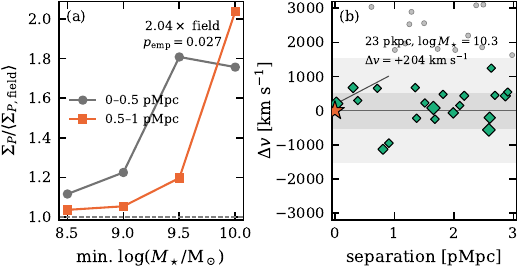}
    \caption{Environment of \src. (a) $P(z)$-weighted surface density of galaxies within $|\Delta z|<0.05$,
    relative to coverage-matched random apertures, against the minimum stellar mass, for the 0--0.5 and
    0.5--1\,pMpc annuli. (b) Projected separation against velocity offset for secure NIRSpec redshifts within
    3\,pMpc, with $\pm500$ and $\pm1500$\,km\,s$^{-1}$ shaded.}
    \label{fig:environment}
\end{figure}
 
We use the DJA photometric catalogue of PRIMER-UDS \citep{Donnan2024} fit in Baker+, in prep to explore the nearby environment of \src. For each galaxy we compute the probability $P_{\rm mem}$ of lying within $|\Delta z|<0.05$ of \src, and the $P(z)$-weighted surface density in annuli, using the photometric $P(z)$ for all galaxies so that the result is not biased by the spectroscopic targeting. 
As \src\ lies close to the edge of the footprint, we compare to 5000 random
apertures per annulus matched in coverage, giving an overdensity and an empirical $p$-value. The overdensity grows with the mass threshold (Fig. \ref{fig:environment}a): galaxies with $\log(M_\star/\Msun)\geq10$ at 0.5--1\,Mpc are $2.04\times$ as common as in the field ($p=0.027$), robust to the
redshift window, whilst lower-mass galaxies show no significant excess. 
This signal is carried by only $\sim$1.1 effective galaxies against 0.6 expected, and the significance is not corrected for the several radii, masses and windows explored. The densest region lies 0.85--1.5\,Mpc from \src. Within 3\,Mpc, 21
galaxies with secure NIRSpec redshifts lie within $\pm1500$\,km\,s$^{-1}$ of \src\ and 13 within
$\pm500$\,km\,s$^{-1}$ (Fig. \ref{fig:environment}b).

Galaxies truly in the structure have on average only 12\% of their $P(z)$ within
$|\Delta z|<0.05$ (from the 40 spectroscopic members), so we divide the excess by this factor, and adopt a bias
of $b\approx5$ for their host haloes \citep{Tinker2010}.
To estimate the mass of the structure, we convert the excess of $\log(M_\star/\Msun)\geq10$ galaxies
into a matter overdensity.  As the line-of-sight extent of the structure is unknown,
we bracket it by placing the excess either within a sphere of the projected radius or across the full redshift
window. On 10--15\,cMpc scales this gives $\delta_{\rm m}\approx0.3$--1 and an enclosed mass of
$\sim10^{14.4\text{--}15.1}$\,\Msun, which collapses by $z=0$ \citep{Mo1996} only if the excess is confined to a sphere of the projected radius.
On smaller scales the excess is dominated by a few galaxies and is non-linear. Growing a halo of $\log(M_{\rm h}/\Msun)\approx12.5$--12.9 with the median accretion history \citep{Fakhouri2010} gives a massive group at $z=0$ ($\sim10^{13.7\text{--}14.1}$\,\Msun), whereas the enclosed mass above would form a rich cluster. With a modest signal, and as $z\approx4$ overdensities are poor predictors of $z=0$ masses \citep{Remus2023}, the descendant of \src\ is therefore uncertain.
 
\begin{figure}
    \centering
    \includegraphics[width=\columnwidth]{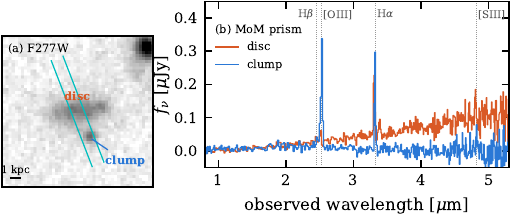}
    \caption{The closest spectroscopic neighbour of \src, 23\,kpc to the north. (a) F277W image ($2\farcs4$
    across) with the MoM-UDS prism shutter, crossing the red disc and a compact clump 0\farcs4 to the
    south-south-west. (b) Prism spectra of the two components.}
    \label{fig:companion}
\end{figure}
 
The closest of the neighbours lies 23\,pkpc north of \src\ at $z=4.055$, with $\log(M_\star/\Msun)\approx10.3$, around a
twelfth of the mass of \src. Its MoM-UDS prism spectrum (Oesch+, in prep) covers two components
(Fig. \ref{fig:companion}). The first is a red, close to edge-on disc with \ha+[NII] (EW$_0\approx200$\,\AA),
weak [OIII] and a Balmer decrement indicating $A_V\sim1$--1.5, and is undetected in ALMA Band 7
(SFR$_{\rm IR}\lesssim70$\,\Msun\,yr$^{-1}$). The second is a compact clump 0\farcs4 (3\,kpc) away, with almost no
continuum and extreme lines ([OIII]$\lambda5007$/H$\beta\approx7$, EW$_0\gtrsim1000$\,\AA), indicating a young,
low-metallicity and dust-free starburst. \src\ is therefore likely to continue to grow through minor mergers.

\begin{table}
\caption{Galaxies with secure NIRSpec redshifts within 3\,pMpc and $\pm1500$\,km\,s$^{-1}$ of \src\
(Fig. \ref{fig:environment}b), ordered by projected separation.}
\label{tab:neighbours}
\centering
\footnotesize
\setlength{\tabcolsep}{2.5pt}
\begin{tabular}{lcccccc}
\hline\hline
ID & RA & Dec & $z_{\rm spec}$ & $r$ & $\Delta v$ & $\log M_\star$ \\
 & (J2000) & (J2000) & & [pMpc] & [km\,s$^{-1}$] & [M$_\odot$] \\
\hline
67248 & 02:17:27.3 & $-$05:06:40 & 4.0548 & 0.02 & $+204$ & 10.3 \\
75129 & 02:17:27.6 & $-$05:06:00 & 4.0628 & 0.31 & $+675$ & 9.4 \\
69279 & 02:17:30.8 & $-$05:06:29 & 4.0564 & 0.38 & $+297$ & 8.4 \\
46864 & 02:17:27.9 & $-$05:08:23 & 4.0624 & 0.71 & $+655$ & 7.8 \\
77519 & 02:17:20.7 & $-$05:05:46 & 4.0323 & 0.81 & $-1132$ & 9.4 \\
79503 & 02:17:19.9 & $-$05:05:39 & 4.0353 & 0.91 & $-953$ & 8.6 \\
35577 & 02:17:34.6 & $-$05:09:20 & 4.0628 & 1.35 & $+675$ & 8.6 \\
53199 & 02:17:39.5 & $-$05:07:51 & 4.0476 & 1.37 & $-223$ & 7.2 \\
71566 & 02:17:41.5 & $-$05:06:18 & 4.0552 & 1.51 & $+225$ & 8.2 \\
19857 & 02:17:27.2 & $-$05:10:37 & 4.0527 & 1.65 & $+79$ & 10.3 \\
50607 & 02:17:42.3 & $-$05:08:04 & 4.0472 & 1.68 & $-251$ & 8.6 \\
41475 & 02:17:42.2 & $-$05:08:51 & 4.0595 & 1.82 & $+479$ & 7.8 \\
50865 & 02:17:45.4 & $-$05:08:03 & 4.0503 & 1.99 & $-66$ & 9.5 \\
59167 & 02:17:07.6 & $-$05:07:22 & 4.0539 & 2.10 & $+150$ & 8.3 \\
21219 & 02:17:13.5 & $-$05:10:30 & 4.0587 & 2.17 & $+435$ & 8.9 \\
70935 & 02:17:02.9 & $-$05:06:22 & 4.0419 & 2.59 & $-564$ & 10.1 \\
71667 & 02:17:02.7 & $-$05:06:18 & 4.0479 & 2.60 & $-207$ & 9.8 \\
17945 & 02:17:08.5 & $-$05:10:46 & 4.0724 & 2.63 & $+1246$ & 8.3 \\
34153 & 02:17:50.3 & $-$05:09:26 & 4.0590 & 2.69 & $+449$ & 7.1 \\
15481 & 02:17:06.4 & $-$05:11:00 & 4.0587 & 2.86 & $+434$ & 9.4 \\
41594 & 02:17:01.2 & $-$05:08:51 & 4.0606 & 2.90 & $+545$ & 8.1 \\
\hline
\end{tabular}
\tablefoot{IDs and stellar masses from the DJA PRIMER-UDS catalogue; redshifts are DJA grade 3. $\Delta v$
is relative to $z_{\rm sys}=4.0514$.}
\end{table}

\section{AGN constraints}
\label{app.agn}
We find no evidence for an energetically important AGN in \src. It is undetected in the X-UDS
\textit{Chandra} data \citep{Kocevski2018}, with $L_{2-10\,\rm keV}<4\times10^{43}$\,erg\,s$^{-1}$ ($3\sigma$), only a
few times that expected from X-ray binaries, ruling out an AGN contributing more than
$\sim$5--10\% of $L_{\rm IR}$ unless it is Compton-thick. As H$\beta$ and [OIII] are not detected, we cannot use
the BPT diagram, but the G395M ratios ($\log$\,[NII]/\ha$=-0.60$ to $-0.25$) and \ha\
equivalent width ($\sim$80\,\AA) are consistent with metal-rich star formation.
It is also undetected at 1.4,GHz in the VLA imaging used by \citet{Dudzeviciute2020} ($S_{1.4}\lesssim30\,\mu$Jy), implying $q_{\rm IR}\gtrsim2$, consistent with star formation and excluding a radio-loud AGN, and the MIRI photometry is well fitted by stars alone. A low-luminosity or Compton-thick AGN in the core cannot be excluded, but is unlikely to contribute significantly.

\begin{acknowledgements}

WMB acknowledges support from DARK via the DARK Fellowship. This work was supported by a research grant (VIL54489) from VILLUM FONDEN. 
FV, KI, and PZ acknowledge support from the Independent Research Fund Denmark (DFF) under grant 3120-00043B. 
W. M. Baker acknowledges use of Claude Opus 5.5 for assistance with coding and language editing.
Some of the data products presented herein were retrieved from the Dawn JWST Archive (DJA). DJA is an initiative of the Cosmic Dawn Center, which is funded by the Danish National Research Foundation under grant DNRF140. 
This work is based on observations made with the NASA/ESA/CSA James Webb Space Telescope, obtained at the Space Telescope Science Institute, which is operated by AURA under NASA contract NAS 5-03127. These observations are associated with programmes GO-1837, 4233 and 7814. This paper makes use of the following ALMA data: ADS/JAO.ALMA \#2013.1.00781.S, \#2015.1.00442.S, \#2015.1.01528.S, \#2016.1.00434.S, \#2021.1.00705.S and \#2023.1.01520.S. ALMA is a partnership of ESO (representing its member states), NSF (USA) and NINS (Japan), together with NRC (Canada), NSTC and ASIAA (Taiwan) and KASI (Republic of Korea), in cooperation with the Republic of Chile. The Joint ALMA Observatory is operated by ESO, AUI/NRAO and NAOJ.



\end{acknowledgements}
 
\end{appendix}

\end{document}